\documentclass[amsmath,amssymb,nofootinbib,showpacs,twocolumn,nofootinbib]{revtex4-1} 
\usepackage{dcolumn}
\usepackage{bm}
\usepackage{graphicx}
\usepackage{epsfig}
\usepackage{color}
\usepackage{wasysym}
\usepackage{natbib}
\usepackage{twoopt}
\usepackage{dashrule}
\usepackage[misc]{ifsym}
\definecolor{slateblue}{rgb}{0.2,0.2,0.6}
\usepackage[breaklinks=true,colorlinks=true,linkcolor=slateblue,citecolor=slateblue,urlcolor=slateblue,pdfauthor={Tomassetti},pdftitle={PropTimes2025}]{hyperref}

\usepackage{xspace}
\newcommand{\ApJ}{ApJ}
\newcommand{\AeA}{A\&A}
\newcommand{\PRL}{PRL}
\newcommand{\PRD}{PRD}
\newcommand{\PRC}{PRC}
\newcommand{\ASR}{ASR}
\newcommand{\JGR}{JGR}
\newcommand{\GCR}{CR\xspace}
\newcommand{\GCRs}{CRs\xspace}

\newcommand{\etal}{et al.}
\newcommand{\ie}{\textit{i.e.}} 
\newcommand{\eg}{\textit{e.g.}} 

\def\citep#1{\cite{#1}}

\begin{document}
\title{Propagation times and energy losses of\\ cosmic
  protons and antiprotons in interplanetary space}

\author{N. Tomassetti$^{\,1\,{\color{slateblue}*}}$, B. Bertucci$^{\,1}$, E. Fiandrini$^{\,1}$, B. Khiali$^{\,2}$}
\address{$^{1}$\,Dipartimento di Fisica e Geologia, Universit{\`a} degli Studi di Perugia, Italy}\email{nicola.tomassetti@unipg.it} 
\address{$^{2}$\,INFN - Sezione di Roma Tor Vergata, Roma, Italy}

\begin{abstract}
  In this paper, we investigate the heliospheric modulation of cosmic rays interplanetary space, focusing on their propagation times and energy losses over the solar cycle.
  To perform the calculations, we employ a data-driven model based on the stochastic method. Our model is calibrated using time-resolved and energy-resolved data
  from several missions including AMS-02, PAMELA, EPHIN/SOHO, BESS, and data from Voyager-1.
  This approach allows us to calculate probability density functions for the propagation time and energy losses of cosmic protons and antiprotons in the heliosphere.
  Furthermore, we explore the temporal evolution of these probabilities over spanning from 1993 to 2018, covering a full 22-year cycle of magnetic polarity which includes
  two solar minima and two magnetic reversals.
  Our calculations are carried out for cosmic protons and antiprotons, enabling us to investigate the role of charge-sign dependent effects in cosmic ray transport.
  These findings provide valuable insights into the physical processes of cosmic-ray propagation in the heliosphere and contribute to a deeper
  understanding of the solar modulation phenomenon.
\end{abstract}
\maketitle
 
\section{Introduction}    
\label{Sec::Introduction} 

Galactic cosmic rays (\GCRs) are excellent astrophysical messengers for probing the high-energy Universe. 
When entering the heliosphere, however, they interact with magnetic fields and solar wind disturbances,
leading to changes in their intensity and energy spectrum. This phenomenon is known as solar modulation of \GCRs.
Understanding solar modulation is crucial for investigating the origins of cosmic particles and antiparticles \citep{Potgieter2013}.
It is also important for evaluating radiation exposure and risks associated with
crewed space missions or orbiting devices \citep{Norbury2018,Cucinotta2015}.
Consequently, considerable efforts are being done directed towards studying the
variations in the \GCR spectrum, both near-Earth and in interplanetary space.
An important aspect of solar modulation is its association with the 11-year solar activity cycle.
The monthly sunspot number (SSN), a primary indicator of solar activity, exhibits an inverse relationship
with the long-term \GCR flux variations \citep{Hathaway2015,Usoskin1998}. 
At the peak of the SSN cycle, when the Sun releases more energy in the form of solar flares and coronal mass ejectionsm
the \GCR flux is minimum. During solar minimum, the \GCR flux attains its maximum level.
Along with the SSN cycle, the Sun shows a 22-year quasiperiodical cycle of magnetic polarity,
with a magnetic field reversal occurring during each solar maximum.
This periodicity is of paramount importance for studying particle drifts
in the interplanetary magnetic field (IMF).

Observationally, the modulation phenomenon is investigated by time- and energy- dependent measurements of \GCRs in space.
Valuable data have been released recently by the AMS-02 experiment on the
International Space Station \citep{Aguilar2018PHeVSTime,Aguilar2021Report,Aguilar2021ProtonDaily},
and by the PAMELA mission in the Resurs-DK1 satellite \citep{Martucci2018,Adriani2013Protons}.
Their data add to earlier measurements conducted, in particular, by
the EPHIN space instrument onboard SOHO \citep{Kuhl2016} and by several BESS balloon flights \citep{Abe2016,Shikaze2007}. 
Measurements of the local interstellar spectra (LIS) of \GCRs have been conducted by the Voyager-1
spacecraft outside the heliosphere \citep{Cummings2016}.
 
From a modeling perspective, an important challenge lies in elucidating the dynamic relationship between
the local \GCR flux and the evolving conditions of the heliospheric plasma.
The main physical mechanisms governing this relatioship include the
diffusion of \GCRs through small-scale IMF irregularities,
convection and adiabatic cooling over the expanding solar wind,
and drift motion across the large-scale IMF \citep{Potgieter2013,Engelbrecht2022}. 
These mechanisms induce significant modifications in the fluxes of \GCRs
in the heliosphere and may explain most of the observed features.
In the recent years, many effective models of solar modulation have been developed
\citep{Manuel2014,Potgieter2017,Tomassetti2018PHeVSTime,Tomassetti2019Numerical,Boschini2018,Corti2019,Wang2020,Song2021,Fiandrini2021,Aslam2023}.

The present work focuses on the propagation times and the energy losses of \GCRs in the heliosphere.
By propagation times here we refer to the time spent by galactic particles entering the heliosphere
to reach the near-Earth environment, \ie, a inner region of heliosphere situated at 1 AU from the Sun.
The energy losses pertains to the process of adiabatic cooling affecting \GCRs
in their travel through the expanding solar wind.
The propagation times and energy losses of \GCRs have been studied in a number of works. 
The gross features were established in early studies by analytical calculations
\citep{Parker1966,Jokipii1967,OGallagher1975}.
More recent works rely on sophisticated numerical models \citep{Strauss2011JGR,Strauss2013,Vogt2022,Moloto2023}. 
These researchers have yielded valuable insights on understanding the timescales of the \GCR transport
in the heliosphere under different conditions of IMF polarity and \GCR energies \citep{Strauss2011JGR}.

In this paper, we explore the temporal evolution of propagation times and energy losses for both \GCR
protons and antiprotons under the real conditions of the past solar cycles.
To achieve this, we calculate the associated probability distributions using a data-driven stochastic model of \GCR modulation.
Our model is globally calibrated using large time-series of \GCR proton measurements as function of kinetic energy and time.
The \GCR data are obtained from space experiments AMS-02, PAMELA, and EPHIN/SOHO, as well as from BESS balloon flights.
The temporal extension of these measurements encompasses the tail of solar cycle 22 and include solar cycles 23 and 24.
We also utilize IMF plasma data from space probes WIND and ACE, along with the Wilcox Solar Observatory.

The rest of this paper is organized as follows. In Sect.\,\ref{Sec::CRPropagation}, we overview
the basic \GCR propagation theory and present our numerical model of solar modulation. 
In Sect.\,\ref{Sec::ModulationParameters}, we outline the procedure for the data-driven determination
of the key model parameters based on measurements of \GCR protons.
In Sect.\,\ref{Sec::Results} we present our results on the \GCR propagation times and the energy loss fractions;
in particular we present calculations for their probability distributions, how they evolve
in different  phases of the solar cycle, how they depend on the \GCR energies and on their charge-sign. 
In Sect.\,\ref{Sec::Conclusions}, we summarize our results and draw the conclusions. 

\section{Theoretical framework} 
\label{Sec::CRPropagation}      

The propagation of high-energy charged particles in
the heliospheric
plasma is generally described by the Parker equation:
\begin{equation}\label{Eq::Parker}
\frac{\partial f}{\partial t} + \nabla \cdot (\vec{V}_{sw} -\textbf{K} \nabla f)-\frac{1}{3}(\nabla \cdot \vec{V}_{sw})\frac{\partial f}{\partial (\ln R)}=0 \,,
\end{equation}
where $f(\vec{r},R,t)$ is the omnidirectional distribution function for a given particle type, $t$ is the time,  $R{\equiv}p/Z$ is the particle rigidity,
\ie, the momentum per charge units. The quantity $\textbf{K}$ is the diffusion-drift tensor and $\vec{V}_{sw}$ is the solar wind velocity.
The differential \GCR flux as function of rigidity is given by $J=R^{2}f$.
There are several approaches to numerically solve Eq.\ref{Eq::Parker} and obtain the differential intensity
for a given \GCR species and at a given location in heliosphere \citep{Potgieter2013}.
Here we use the calculation framework presented in details in \citet{Fiandrini2021,Tomassetti2023MFP},
which is briefly recapitulated in the following sections.
From Eq.\ref{Eq::Parker}, it can be seen that the distribution function resulting from the propagation of \GCRs in
the helisphere is time-dependent, and many characteristic parameters of the ambient medium are subjected to temporal variation.
However, in order to study the essential features of \GCR fluxes averaged over Bartels rotation lasting 27 days, or on longer time periods,
the steady-state approximation ($\partial f/\partial t=0$) is often considered a reasonable working assumption. 
In such a description, the long-term modulation of \GCRs is described as a continuous series of steady-state solutions
where the effective status of the heliospheric plasma during the \GCR propagation is defined in a suitable way.
Such an approach is usually justified by the different timescales between \GCR transport in the heliosphere and the changing solar activity, 
which will be assessed and discussed in this paper. 
In order to carry out a realistic description of the proton modulation, we need first to clarify the three principal ingredients of this study:
(i) the physical characteristic of the modulation region;
(ii) the diffusion and drift coefficients that enter the transport equation;
(iii) the input LIS for cosmic protons and antiprotons. 
In the following subsections we give a brief description of them.

\subsection{Key parameters of the modulation domain} 
\label{Sec::Heliosphere}                             

In our framework, the modulation region is the entire heliosphere.
Its spatial extent is defined by the heliopause and it is modeled as a spherical surface of radius $r_{b}=122$\,AU.  
The Earth position is located in the equatorial plane at helioradius $r=r_{0}=1$\,AU.
The solar wind flows out radially from the Sun.
During solar minimum, the wind is fast in the polar region of the heliosphere (${7-800}$\,km\,s$^{-1}$)
and slow in equatorial plane (${3-400}$\,km\,s$^{-1}$).
During solar maximum, the wind shows no clear latitudinal dependence and
its average speed is ${4-500}$\,km\,$s^{-1}$ at all latitudes \citep{Heber2006}. 
The wind speed remains remarkably constant with radius, up to the termination shock located at $r\approx$\,85\,AU.
Beyond the termination shock, the wind slows down by a factor $1/S$, where $S{\approx}2.5$ is the shock compression ratio.
All these features are captured by a parametric function $V_{sw}(t,r,\theta)$ of
time, radial and latitudinal coordinates. 
Like in our past work \cite{Fiandrini2021}, we assume a factorized form for the wind speed profile:
\begin{equation}\label{Eq::SW}
  V_{sw}(t,r,\theta)= V_{0} \times g(t,\theta) \times f(r) \,,
\end{equation}
where we adopt the parametric expression given in Ref.\,\cite{Potgieter2014}. 
The termination shock is accounted in the radial part $f(r)$. The temporal dependence of $V_{sw}$ consists in the
changing angular extent of the slow wind region, which we have related to the IMF tilt angle \citep{Fiandrini2021}.
The normalization is set by the parameter $V_ {0}\cong 450\,km\,s^{-1}$. 
To model the large-scale structure of the IMF we use the modified Parker model:
\begin{equation}\label{hmf}
  B=B_{0}(t)\left(\frac{r_0}{r}\right)^{2}\bigg\{1+\tan^2\psi +\bigg(\frac{r\delta(\theta)}{r_{\odot}}\bigg)^2 \bigg\}^{1/2} \,,
\end{equation}
where $B_{0}$ is the \emph{local} IMF value at $r=r_{0}=1$\,AU, which is continuously measured \emph{in situ} by spacecraft.
The quantity $\psi$ is the so-called winding angle between the IMF and the radial direction $(\tan^{2}\psi\approx{\Omega}r/V)$,
and the $\delta(\theta)$ function is a high-latitude correction \citep{JokipiiKota1989}.
This description also includes the heliospheric current sheet, the thin layer where IMF polarity changes from North to South.
The current sheet is wavy due to the tilt angle $\alpha$ between the solar magnetic axis and its rotational axis.
The tilt angle is time-dependent, varying from $\alpha{\sim}\,5^\circ$ during solar minimum
to $\alpha{\gtrsim}\,70^\circ$ during solar maximum.
The IMF polarity, $A$, is defined by the sign of $\vec{B}$ in proximity of the Sun.
Polarity is positve when $\vec{B}$ is directed outwards in its northern hemisphere
and inward in the southern hemisphere. Polarity reversals occur during every solar maximum. 
The \GCR transport is influenced by changes in polarity, as the drift speed
depends on the sign product $q{\times}A$ between charge and polarity.

\subsection{\GCR diffusion and drift} 
\label{Sec::DiffusionAndDrift}       

Diffusion and drift processes are captured by the tensor ${\bf{K}}$ appearing in the third term of Eq.\,\ref{Eq::Parker}.
\begin{equation}\label{k}
{\bf K}= {\bf K^{S}} + {\bf K^{A}} =
\begin{bmatrix}
 K_{r\perp} & -K_A & 0\\
K_A & K_{\theta\perp} & 0\\
0 & 0 & K_{\parallel}
\end{bmatrix}
\end{equation}
where ${\bf K^{S}}$ and ${\bf K^{A}}$ are the symmetric and antisymmetric parts of the tensor, respectively, describing diffusion and drift.
The symmetric part includes the coefficient $K_{\parallel}$, associated with parallel diffusion along the field direction,
and the perpendicular coefficients $ K_{r\perp}$ and $K_{\theta\perp}$ describing transverse radial and transverse polar diffusion, respectively. 
All coefficients can also be expressed in terms of mean free path $\lambda$, by means of relations
such as $K_{\parallel}=\beta c\lambda_{\parallel}/3$, where $\beta=v/c$.
Like in \citep{Fiandrini2021}, and following earlier works
\cite{Potgieter2014,Zank1998}, we model the spatial and rigidity dependence of the \GCR parallel diffusion with the following formula:
\begin{equation}\label{Eq::KvsR}
 K_{\parallel}=K_0\frac{\beta}{3}\frac{(R/R_0)^a}{(B/B_0)}\bigg[\frac{(R/R_0)^h+(R_k/R_0)^h}{1+(R_k/R_0)^h}\bigg]^{\frac{b-a}{h}} \,,
\end{equation}
where $K_{0}$ is of the order of $10^{23}$\,cm$^2s^{-1}$ and $R_0{\equiv}1\,GV$ to sets the rigidity units.
The rigidity dependence is a broken power-law function describing the two diffusion regimes \cite{Teufel2002}.
The spectral indices $a$ and $b$ set the log-slopes below and above the characteristic rigidity $R_{k}=3\,GV$.
The latter represents the scale rigidity value where the GCR gyroradius matches the correlation length of the IMF power spectrum. 
The parameter $h{\equiv}0.01$ determines the smoothness of the transition.
The perpendicular coefficients are determined by the scaling assumptions
$K_{\perp, r/\theta}\cong\xi_{r/\theta}K_{\parallel}$ with $\xi_{r/\theta}{\cong}0.02$ \citep{Giacalone1999}. 
The overall spatial dependence of \GCR diffusion is determined by the IMF model via $B$.
Its temporal dependence is therefore capture by the free parameters $K_{0}$, $a$ and $b$, that are determined by \GCR data.
The antisymmetric part of the tensor is specified by the drift coefficient:
\begin{equation}\label{Eq::KA}
 K_{A}=\frac{K^0_A}{3}\frac{\beta R}{B}\frac{(R/R_A)^2}{1+(R/R_A)^2} \,,
\end{equation}
where $R_{A}\cong\,0.5\,GV$ is the rigidity below which the drift is reduced due to scattering.
The normalization factor $K^{0}_{A}$ is set to unity by default. In other works, low $K^{0}_{A}$ values are often used to reduce drift during the reversal phase.
In our model, however, the flux during reversal is modeled as a superposition of fluxes with opposite polarities \citep{Fiandrini2021,Jiang2023},
and thus we do not need to suppress $K_{A}$ manually.
From Eq.\,\ref{Eq::KA}, the average drift velocity is given by $\langle V_{D} \rangle = \nabla\times{K_{A}}$.

\subsection{Local interstellar spectra} 
\label{Sec::LIS}                        

The LIS of \GCRs beyond the heliopause is an important factor for solar modulation models.
In this study, we focus on \GCR protons and antiprotons. To determine their LIS, a dedicated modeling effort is required.
Here, we employ the results of a two-halo model of \GCR propagation in the
Galaxy \citep{Tomassetti2012Hardening,Tomassetti2015TwoHalo,Feng2016}.
In this model, protons and other primary nuclei accelerated in the shocks of supernova remnants
and then injected in the interstellar space.
The primary acceleration spectrum of \GCRs is described by a source term of the type
$S_{\rm p}\propto(R/{\rm GV})^{-\nu}$,
with index $\nu=$\,2.28\,$\pm$\,0.12 for $Z=1$ and index $\nu=$\,2.35\,$\pm$\,0.13 for all $Z>1$ nuclei.
The production of secondary nuclei
is then calculated by a source term  $S_{\rm s}= \sum_{\rm h} \Gamma_{h{\rightarrow}s}^{\rm sp} N_{\rm h}$.
This term describes the $h{\rightarrow}s$ fragmentation of $h$-type species of density $N_{\rm{h}}$ into $s$-type nuclei at
rate $\Gamma_{h{\rightarrow}s}(E) = n v_{h} \sigma_{h{\rightarrow}s}(E)$, where $n$ is the gas density
and $\sigma_{h{\rightarrow}s}$ is the cross-sections for the $h{\rightarrow}s$ process.
The calculations depends on many cross-sections, that have been estimated in earlier works \citep{Tomassetti2017BCUnc,Tomassetti2015XS}.
For the production of secondary nuclei, it is assumed that kinetic energy per nucleon is
maintained (\ie, $E_{h}\cong E_{s}\equiv E$), while for antiprotons this is not the case.
The $\bar{p}$ source term is of the type:
\begin{equation}
  S_{\bar{p}}(E) = \sum_{h} \int \Gamma_{h{\rightarrow}\bar{p}}(E, E_{h})^{\rm sp} N_{\rm h} dE_{h} \,,
\end{equation}
where the production rate depends on the differential cross-section $d\sigma_{h{\rightarrow}\bar{p}}/dE_{h}$ \citep{Feng2016}.
Both \GCR acceleration and secondary production occur in the galactic disk, where sources and matter are located. 
Away from the disk, the \GCR propagation is described by a two-halo diffusion model.
The \GCR diffusion coefficient is $D \equiv \beta D_{0}(R/GV)^{\delta_{i/o}}$,
where we assume two diffusion regimes: a shallow diffusion for the near-disk region
with index ${\delta_{i}}$, occurring within distance $z{\leq}\,\xi\,{L}$ from the disk (inner halo),
and a faster diffusion in the extended halo with index ${\delta_{o}}\equiv\delta_{\i}+\Delta$ (outer halo).
Other processes such as ionization losses, destruction processes are also accounted.
Based on previous works, we found $\delta_{i}\,=\,$0.18$\,\pm$\,0.05 and  $D_{0}/L\,=$\,0.01\,$\pm\,$0.002\,kpc/Myr,
$\Delta\,=\,0.55\,\pm$\,0.11, and $\xi = 0.12\,\pm$\,0.03.
%
\begin{figure}
 \centering
 \centering\includegraphics[width=0.44\textwidth]{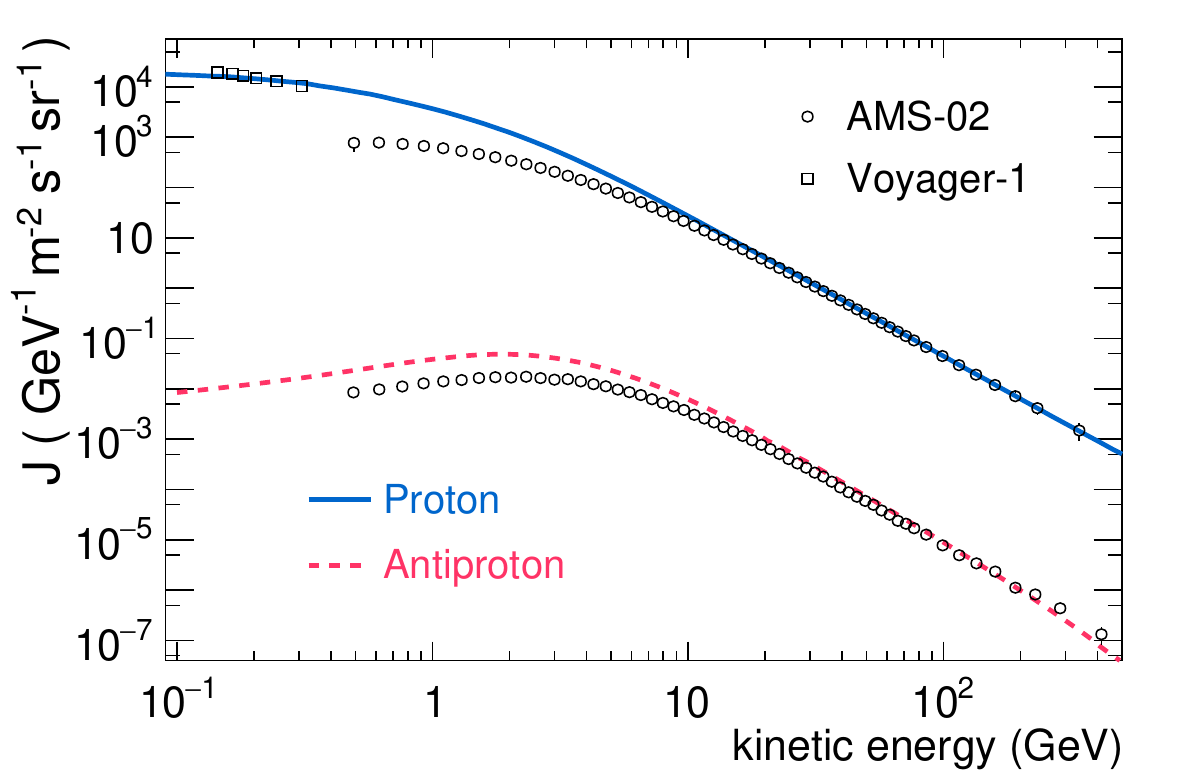}
 \caption{Local interstellar spectra of \GCR protons and antiprotons used as input boundary condition for the modulation along with the estimated uncertainty band.
   The data from Voyager-1 in interstellar space and from AMS-02 in low Earth orbit \citep{Cummings2016,Aguilar2021Report}.}
   \label{Fig::ccLISProtonAntiproton}
\end{figure}
%
The key model parameters are constrained using LIS data from Voyager-1 \citep{Cummings2016},
high-energy measurements on primary nuclei (H-He-C-O) 
and secondary to primary ratios from AMS-02 \citep{Aguilar2021Report}
The calculations are described in details in earlier works \citep{Tomassetti2015TwoHalo,Tomassetti2017BCUnc,Feng2016}. 
The resulting LIS fluxes of \GCR protons and antiprotons are evaluated at near-Earth position,
in galactic cylindrical coordinates $z_{\odot}\,\cong$\,0 of height and $r_{\odot}\cong$\,8.3\,kpc.
The resulting LIS fluxes as a function of kinetic energy are is shown in
Fig.\,\ref{Fig::ccLISProtonAntiproton} in comparison with the data.

\section{Methodology}             
\label{Sec::ModulationParameters} 

\subsection{The stochastic method}  
\label{Sec::SDE}                    

Our calculations are based on the so-called stochastic differental equation (SDE) method.
It consists in transforming the Parker equation into a set of SDEs that can be integrated using Monte-Carlo techniques.
The solution of Eq.\,\ref{Eq::Parker} is then obtained by sampling.
The SDE method is implemented in several solar modulation codes and in the recent years, with the increase
of computing power and resources, is has become very popular \citep{Yamada1998,Zhang1999,FlorinskiPogorelov2009,Pei2010,Kopp2010,Kappl2016,Boschini2018}.
A nice review on this subject is provided in \citet{Strauss2017}. 
For a stochastic process, the stochastic equation can be recasted in the form:
\begin{equation}
  \delta\vec{r} = (\nabla\cdot{\bf K^{S}} - \vec{V}) \delta{t} + {\bf S}\cdot \delta\vec{W} \,,
\end{equation}
where $\vec{V}=\vec{V}_{sw} + \langle \vec{V}_{d} \rangle$ is the total velocity of the particle,
${\bf S}$ is a matrix such that ${\bf S}\cdot{\bf S}=2{\bf K^{S}}$, and
$\delta\vec{W}$ is a multidimensional Wiener process related to a normalized gaussian distribution.
The corresponding change of kinetic energy $E$ in the time ${\delta}t$, for a \GCR proton with mass $m_{p}$ is given by:
\begin{equation}
{\delta}E = \left( \frac{ 2 V_{sw} }{3r}\right) \frac{E^{2} + 2 m_{p} E}{E + m_{p}}  {\delta}t
\end{equation}
The SDE method returns the normalized transition function $G(E,E_{0})$ 
between an initial energy $E_{0}$ at the boundary $r_{p}$ and a final energy $E$ at the Earth's position.
The near-Earth phase space density $f(E)$ is therefore given by:
\begin{equation}
f(E) = \int G(E,E_{0}) f^{\rm LIS}(E_{0})  dE_{0} \,,
\end{equation}
where $f^{\rm LIS}(E_{0})$ the space density corresponding to the LIS flux $J^{\rm LIS}(E_{0})$.

\begin{figure}
 \centering
 \centering\includegraphics[width=0.44\textwidth]{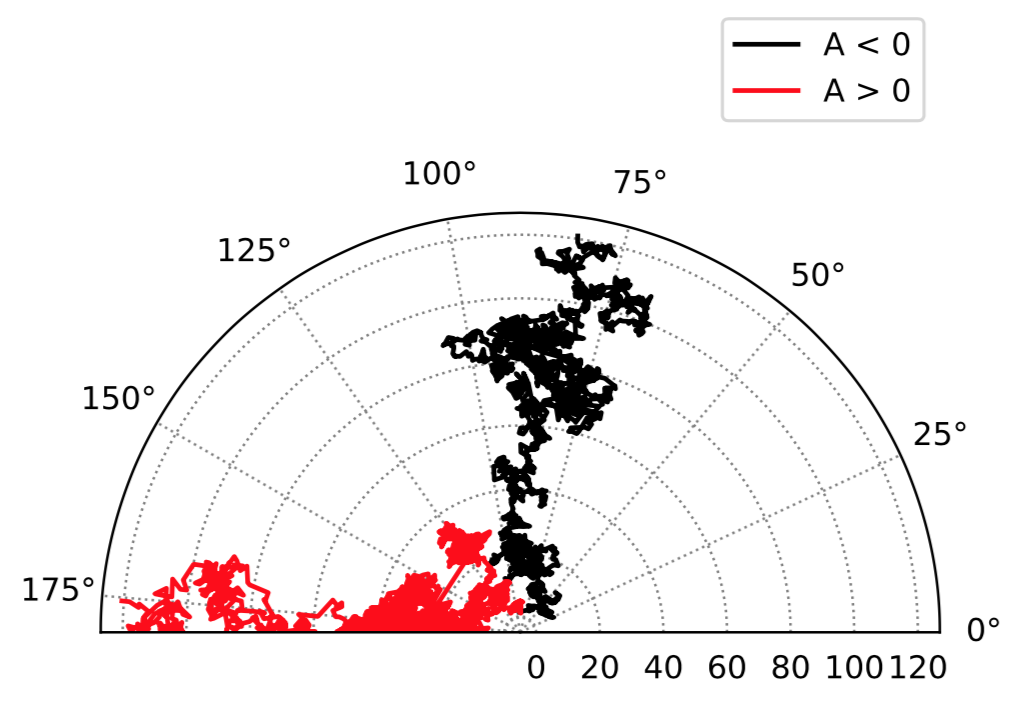}
 \caption{Two simulated trajectories for a 1 GeV proton in polarity phase $A<0$ (black) and $A>0$ (red) represented 
 in a polar diagram of the radial and colatitudinal coordinates.
 }
  \label{Fig::ccTrajectory}
\end{figure}
%
The integration of the equations, in practice, requires the simulation of a large number of pseudo-trajectories.
An example of two typical trajectories is illustrated in  Fig.\,\ref{Fig::ccTrajectory} 
where the radial and colatitudinal coordinates are represented in a polar diagram.
The figure shows the brownian motion of 1\,GeV protons traveling from the heliopause (placed at $r=120$\,AU) to the Earth's vicinity ($r=1$\,AU).
Our IMF model is azimuthally symmetric. 
The two trajectories are related to opposite situations of IMF polarity that will be discussed in Sect.\,\ref{Sec::Results}.
To efficiently sample the steady-state solution in the Earth's vicinity, the SDE generated trajectories were propagated backward in time,
starting from Earth and until they to the heliospheric boundary \citep{Yamada1998,FlorinskiPogorelov2009,Kopp2010}. 
For each pseudo-particle $k$, the method utilizes a time-step incremental scheme, $t_{i}^{k}=t^{k}_{i-1}+{{\delta}t}_{i}$,
and we can determine the total duration of the trajectory as $\tau^{k}{\equiv}\sum_{i} \delta t^{k}_{i}$.
Such a trajectory duration, calculated from the particle source to the observational point, is simply called \emph{propagation time}. 
With a large number $N$ of simulated trajectories, the expectation value of the propagation time can be calculated as:
\begin{equation}\label{Eq::AvgTime}
\langle \tau \rangle = \frac{\int_{0}^{\infty} t f(E,r,t) dt }{ \int_{0}^{\infty}f(E,r,t)dt} \approx \frac{1}{N}\sum_{k=1}^{N} {\tau}^{k} N^{k} \,,
\end{equation}
where $N^{k}$ is the number of pseudo-particles for which the trajectory time falls in a bin around ${\tau}^{k}$.
In some papers, propagation time or its expectation value are called residence time of \GCR in the heliosphere.
It is important to note, however, that the quantity $\langle \tau \rangle$ 
is calculated specifically for those \GCR trajectories that reach Earth's position. 
The kinetic energy loss $E_{k}$ associated to a trajectory can be computed in a similar way, together with its expectation value $\langle E \rangle$. 
Since energy changes account only for adiabatic cooling, the simulated \GCR particles always end with a net energy loss: ${\Delta}E_{k}=E_{0}-E_{k}>0$.
It should also be noted, however, that the backward propagation scheme leads to an increase of the pseudo-particle energy during the simulation.
This is simply due to the time-reversed adiabatic processes.
In the following we will refer to the \emph{energy loss fraction} $\varepsilon \equiv {\Delta}E_{k}/E_{0}$ and its expectation value $\langle\varepsilon\rangle$.
For a given pseudo-proton trajectory, the loss fraction is calculated relatively to the proton initial energy $E_{0}$. 
The SDE method is well suited to calculate $\tau$ and $\varepsilon$. Its applicability is known since its
first formulations \citep{Yamada1998,Zhang1999}.
As we will discuss in Sect.\,\ref{Sec::Results}, propagation times and energy losses of \GCRs depend on several factors
related to the conditions of the background plasma and to the particle properties.

\subsection{Modeling the time dependence} 

In our framework, near-Earth \GCR fluxes are calculated in terms of a time series of steady-state solutions
of Eq.\,\ref{Eq::Parker}, with a resulting proton flux $J_{\rm p}(t,E)$ associated with a time series of input parameters $\vec{q}=\vec{q}(t)$. 
The input parameters can be divided in two classes. On one side, the \emph{heliospheric parameters}
that characterizes the variable conditions of the Sun and the surrounding plasma, \ie, the background medium where the \GCR propagation takes place.
On the other side, the \emph{transport parameters}
that describe the physical processes of \GCR propagation through the heliosphere.
In our model, we have identified a basic set of six time-dependent parameters that
best capture the \GCR modulation phenomenon \citep{Fiandrini2021}.
The heliospheric  parameters are:
the tilt angle of the heliospheric current sheet $\alpha=\alpha(t)$, the intensity of the local IMF $B_{0}=B_{0}(t)$, and its polarity $A=A(t)$.
The transport parameters are:
the normalization coefficient of the parallel diffusion tensor $K_{0}=K_{0}(t)$, and the two spectral
indices $a=a(t)$ and $b=b(t)$, describing the rigidity dependence of
\GCR diffusion below and above the break $R_{k}$, as seen in Eq.\,\ref{Eq::KvsR}. 
For convenience, all parameters are expressed here as continuous functions of time $t$. In practice, their value
has been evaluated from observations and eventually interpolated in defined time-grid of epochs.

\subsection{Determination of the heliospheric parameters} 
\label{Sec::HeliosphericParameters}                       

\begin{figure*}[ht]
 \centering
 \centering\includegraphics[width=0.8\textwidth]{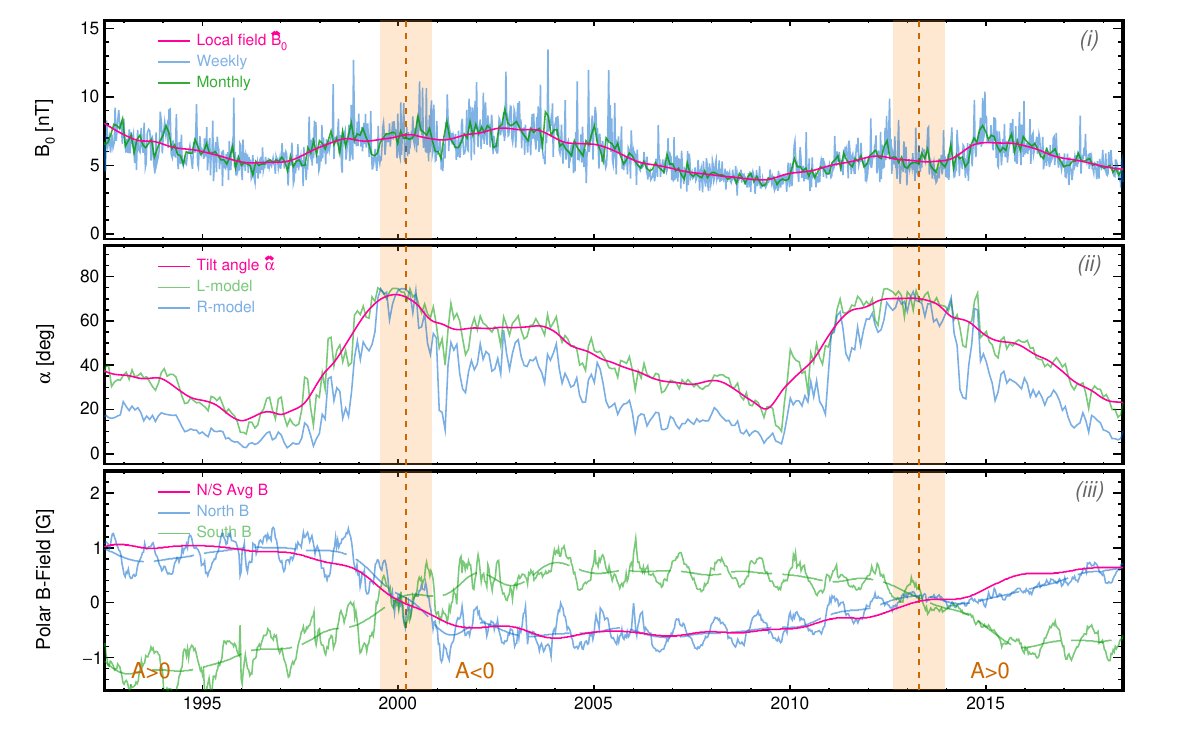}
 \caption{Reconstruction of the heliospheric parameters as function of time:
   (i) local IMF value measured on daily (blue) and weekly basis (green), along with its smoothed reconstruction (pink thick line);
   (ii) tilt angle of the heliospheric current sheet reconstructed on 27-day basis within L-model (green) and the R-model (blue);
   (iii) outward component of the polar magnetic field of the Sun in the north (blue) and south (green) hemisphere, measured on 10-day basis, along with their
   yearly smoothed values (long dashed lines) and their N/S average (pink solid line).
   The sign of the N/S average sets the magnetic polarity $A$.
   The data are from the space probes WIND and ACE and from the Wilcox Solar Observatory \citep{Smith1998,Hoeksema1995,Sun2015}.
}
 \label{Fig::HeliosphericParameters}
\end{figure*}

The temporal dependence of the heliospheric parameters $\{\alpha, B_{0}, A\}$ is determined from observations.
Time-series of tilt angle and polarity are continuously provided by the Wilcox Solar Observatory on 10-day or 27-day basis, respectively.
Measurements of the local IMF $B_{0}$ are done \emph{in-situ} on daily basis by spacecraft WIND and ACE on $L1$ orbit \citep{Smith1998}.
From these data, we build smoothed and continuous functions of time ${\alpha}(t)$, ${B}_{0}(t)$ and ${A}(t)$
by following the procedure of \citet{Fiandrini2021}, but with a wider time coverage.
The method is based on the calculation of the backward moving average over a moving interval $[t-\tau, t]$.
The extent of the interval $\tau$ ranges between 8 and 16 months, depending on the angular
profile of $V_{sw}$ at a given epoch.
For the tilt angle $\alpha$ a 16-month interval is used, because the current sheet lies entirely
within the slow wind region, while for $B_{0}$ it depends on the latitudinal profile of the wind speed at a given epoch.
The results are shown in Fig.\,\ref{Fig::HeliosphericParameters} between 1993 and 2018.
In the figure, the backward moving average functions are shown as tick solid lines, along with the measurements.
We note that the considered time range spans over two solar cycles.
It includes the magnetic reversal of cylce 23, when polarity changed from $A>0$ to $A<0$,
and the latest reversal of cycle 24, where polarity changed from $A<0$ to $A>0$.
Note that, from the figure, the polarity $A$ is given by the sign
of the N/S averaged and polar field, \ie, the pink line of panel (iii).

\subsection{Determination of the transport parameters} 
\label{Sec::TransportParameters}                       

\begin{figure*}[ht]
 \centering
 \centering\includegraphics[width=0.8\textwidth]{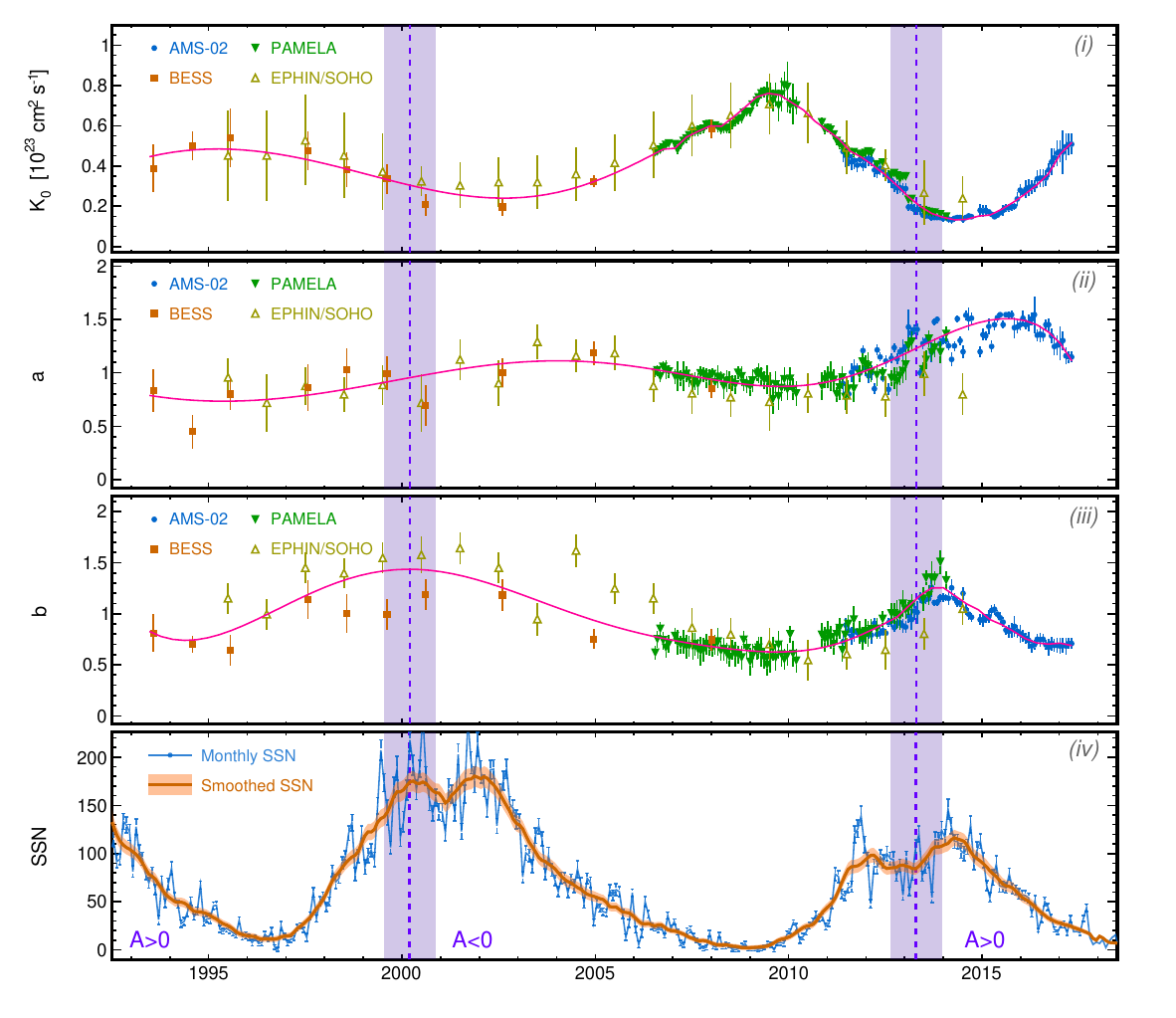}
 \caption{
   Best-fit results for the \GCR propagation parameters $K_{0}$, $a$, and $b$ obtained with 
   proton flux measurements from AMS-02, PAMELA, EPHIN/SOHO and BESS, from panel (i) to (iii).
   The pink solid line is a simple parametric model that connects smoothly the best-fit results.
   In panel (iv), monthly SSN and their 13-month smoothed values.
   The vertical band indicates the magnetic reversal epochs.} 
 \label{Fig::ccTransportParameters}
\end{figure*}

The time-dependent diffusion parameters of the model $K_{0}(t)$,  $a(t)$ and $b(t)$ have been determined
from \GCR proton measurements by a standard procedure of statistical inference.
While the method is similar to that of our previous paper \citep{Fiandrini2021}, here we used a richer sets of data covering the period between 1993 and 2018.
We used monthly-resolved data on the \GCR proton flux collected by the AMS-02 experiment in the International Space Station in 2011-2018,
consisting into a set of 79 Bartels-averaged fluxes measured between 2011 and 2018 \citep{Aguilar2018PHeVSTime} 
in the kinetic energy range from 0.4\,GeV to 60\,GeV.
We also included the 47+36 Bartels-averaged proton fluxes measured by the PAMELA experiment onboard Resurs-DK1 from 2006 to 2014
at kinetic energies from 80\,MeV to 50\,GeV \,\citep{Martucci2018,Adriani2013Protons}.
For past epochs, we used 20 yearly-averaged fluxes from the EPHIN/SOHO probe observed between 1995 to 2015 
at kinetic energies from 0.29\,GeV to 1.6\,GeV. 
and 9 measurements of the proton flux made by the BESS balloon flights between 1993 and 2005 
at kinetic energies from 0.2\,GeV to 20\,GeV.
The whole sample corresponds to a total of about 15,000 data points
collected over a time span of nearly 25\,years.
To sample the six dimensional parameter space, a discrete grid of the vector
$\vec{q}$=\{$\alpha$, $B_0$, $A$, $K_0$,  $a$, $b$\} was built.
For all the 938,400 nodes of the grid, calculations of the \GCR proton flux $J_{m}(E, \vec{q})$ were carried out and
evaluated over 120 values of kinetic energy, ranging from 20\,MeV to 200\,GeV with log-uniform step.
Then, for each set of measurements $J_{d}(E,t)$, representing time-series of energy-binned fluxes, 
the following $\chi^{2}$ estimator was evaluated:
\begin{equation}
\label{Eq::Chi2}
\chi^{2}(\vec{q} )= \sum_{i}  \frac{\left[ J_{d}(E_{i},t) - J_{m}(E_{i}, \vec{q}) \right]^{2} }{\sigma^2(E_{i},t)}
\end{equation} 
Using a 5-dimensional multilinear interpolation technique, the $\chi^{2}$ estimator 
is built such to be a continuous
function of the parameters $\vec{q}$, except for the dichotomous variable $A$.
From the $\chi^{2}(\vec{q})$ function, the best-fit transport parameters $\{{K}_{0},{a},{b}\}$ and their
corresponding uncertainties have been determined at any epoch by standard minimization techniques.
In this procedure, the minimization involves only three free parameters out of six. The remaining
heliospheric parameters were considered as fixed inputs, as they are
determined by the epoch $t$ with reconstruction presented in Sect.\,\ref{Sec::HeliosphericParameters}. 

Our results for the transport parameters  are summarized in Fig.\,\ref{Fig::ccTransportParameters},
showing their best-fit values and uncertainties obtained for the various datasets.
The corresponding SSN cycle is also shown, in the bottom panel.
In the figure, the vertical dashed lines and the shaded bands around them represent the reversal phases. 
It can be seen that the $K_{0}$ parameter shows a distinct temporal dependence, in anti-correlation with the SSN cycle.
\GCRs are subjected to a slower diffusion during epochs of solar minimum, and faster during solar maximum.   
The temporal dependence of $a$ and $b$ parameters is less pronounced, although some variability
can be observed for these parameters, in connection with the solar cycle.
This may be suggestive of variability in heliospheric turbulence over the solar cycle \citep{Fiandrini2021,Vaisanen2019}.
In the figure, the solid line represents a simple parametric model that connects smoothly the best-fit results.
With these lines, we describe the evolution of $\hat{K}_{0}(t)$, $\hat{a}(t)$, and $\hat{b}(t)$ as continuous function of time.
This enables us to specify the whole rigidity and time dependence of the \GCR diffusion coefficients $K_{\parallel,\perp}(t, R)$
over the two considered solar cycles (and similarly, their associated  mean free path $\lambda_{\parallel,\perp}(t, R)$).
Once the transport and heliospheric parameters are established, our measurement-validated model can be used
for investigating the propagation times and energy losses of \GCRs in the heliosphere.

\section{Results and discussions}    
\label{Sec::Results}                 

\begin{figure}[ht]
 \centering
 \centering\includegraphics[width=0.47\textwidth]{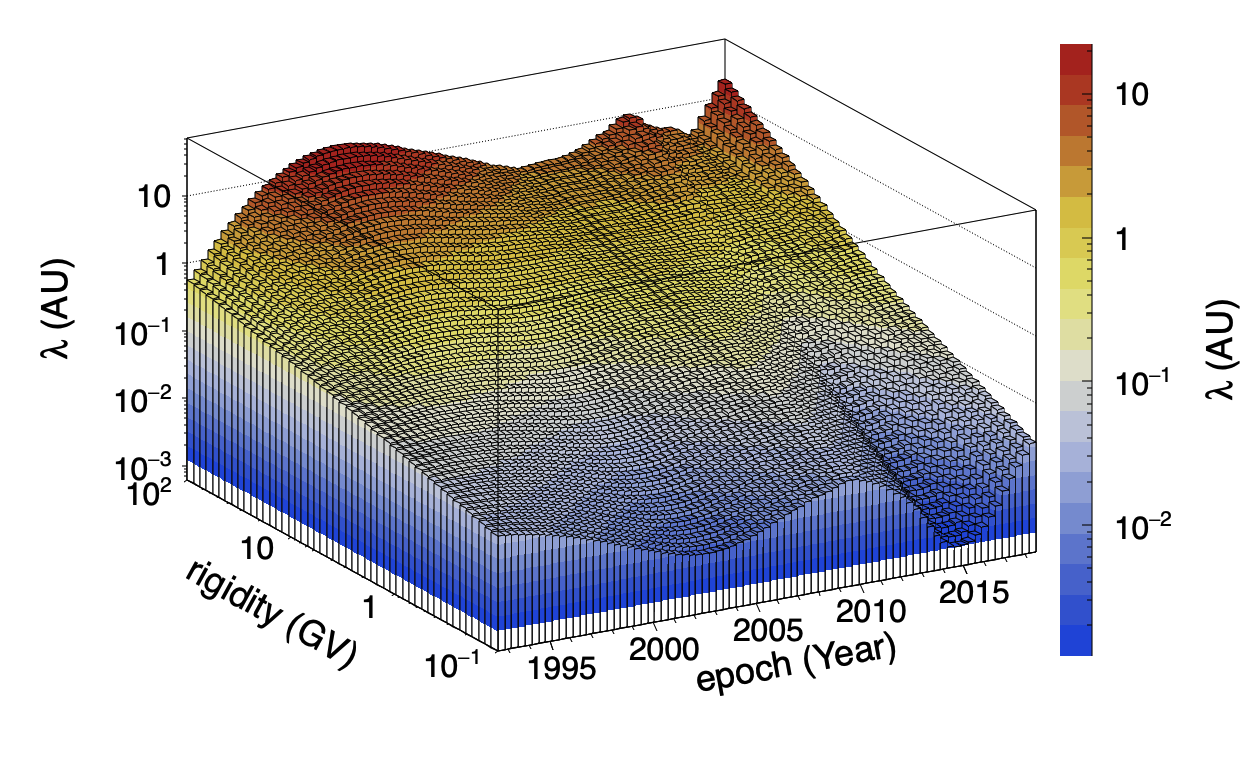}
 \qquad
 \centering\includegraphics[width=0.45\textwidth]{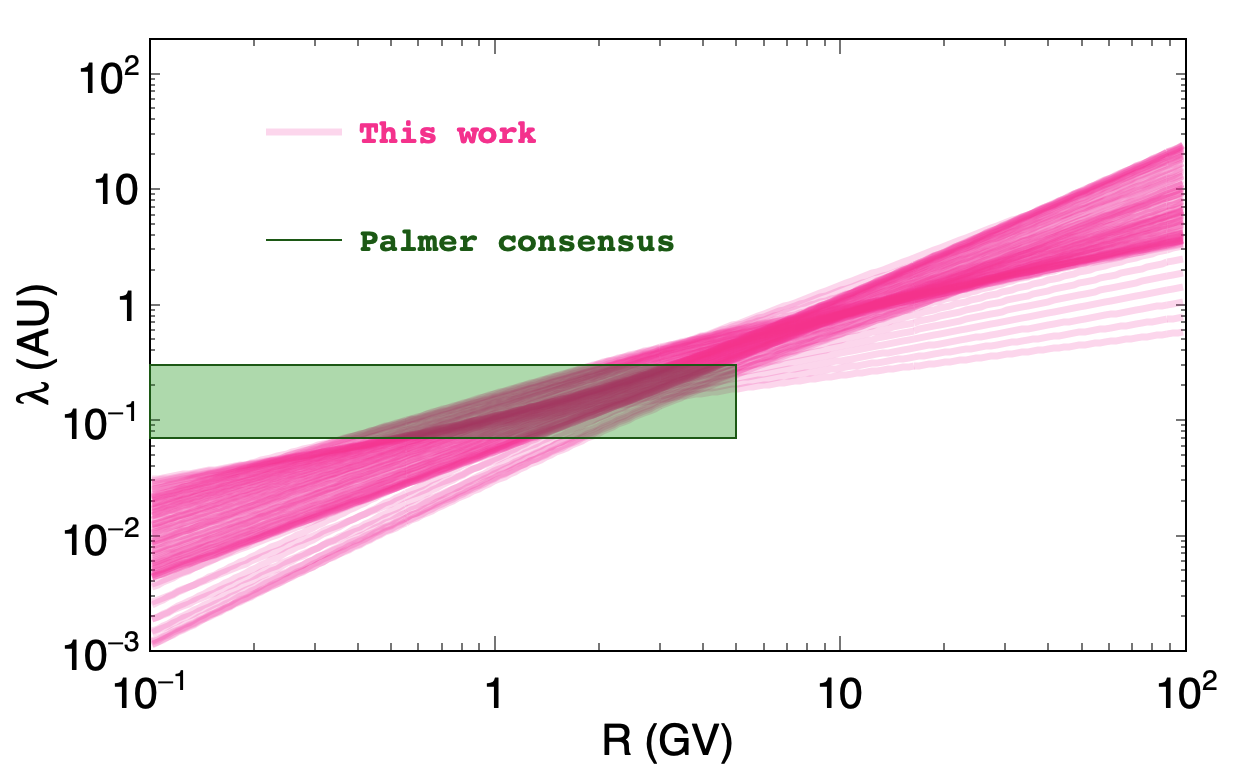}
 \caption{
   Top: mean free path of \GCR parallel scattering $\lambda_{\parallel}(R,t)$ obtained from our fits as function of rigidity and time. 
   Bottom: envelope of the many $\lambda_\parallel$-functions inferred from the data (pink shaded band) compared
   with the ``Palmer consensus'' (green shaded box) \citep{Palmer1982}.} 
 \label{Fig::MFP}
\end{figure}

With the global parametric model of \GCR transport parameters discussed in Sect.\,\ref{Sec::ModulationParameters},
we can now calculate the flux modulation of \GCRs during any epoch in the solar cycle.
From the best-fit parameters of Fig.\,\ref{Fig::ccTransportParameters}, using Eq.\,\ref{Eq::KvsR},
the mean free path for parallel diffusion $\lambda_{\parallel}=\lambda_{\parallel}(R,t)$ can be readily calculated.
The mean free path is shown in Fig.\,\ref{Fig::MFP} (left) as a function of rigidity and time.
From the figure, it can be seen that the diffusion mean free path is subjected to changes in both normalization and slope.
In the right, the envelope of the many $\lambda_\parallel$-functions determined from the data (pink shaded band)
is compared with the so-called ``Palmer consensus'' (green shaded box). The latter is based on past observations between 50\,MeV and 5\,GeV \citep{Palmer1982}.

Since we have adopted a trajectory-based approach, we can compute the distribution
of \GCR propagation times and energy losses associated with the simulated trajectories.
As discussed in several prior studies \citep{Kappl2016,Potgieter2021,Fiandrini2021,Aslam2023,Jiang2023},
the \GCR modulation is significantly influenced by magnetic reversal, which is a remarkable manifestation of the 22-year solar polarity cycle.
Furthermore, \GCRs are also affected by IMF intensity, its large-scale structure, its turbulence levels, which are typically
captured by heliospheric proxies such as SSN, IMF tilt angle, its intensity $B_{0}$ or its variations $\delta B_{0}$.
Based on these considerations, we have identified four reference epochs corresponding to distinct phases of the solar polarity cycle:
(a) January 2009, corresponding to the solar minimum under negative IMF polarity (across cycles 22 and 23, $A<0$).
(b) January 2002, corresponding to the solar maximum and negative IMF polarity (cycle 23, following the reversal from $A>0$ to $A<0$).
(c) March 1996, corresponding to the solar minimum and positive IMF polarity (across cycles 22 and 23, $A>0$).
(d) March 2014, corresponding to the solar maximum and positive IMF polarity (cycle 24, following the reversal from $A<0$ to $A>0$).
It should be noted that magnetic reversal occurs during the peak of solar maximum; 
therefore, epochs (b) and (d) have been selected slightly after reversal, when the magnetic polarity can be clearly defined \citep{Sun2015}.
The results presented in the following sections focus on these reference epochs.

\subsection{Propagation times} 
\label{Sec::PropagationTimes}  

\begin{figure*}
 \centering
 \centering\includegraphics[width=0.8\textwidth]{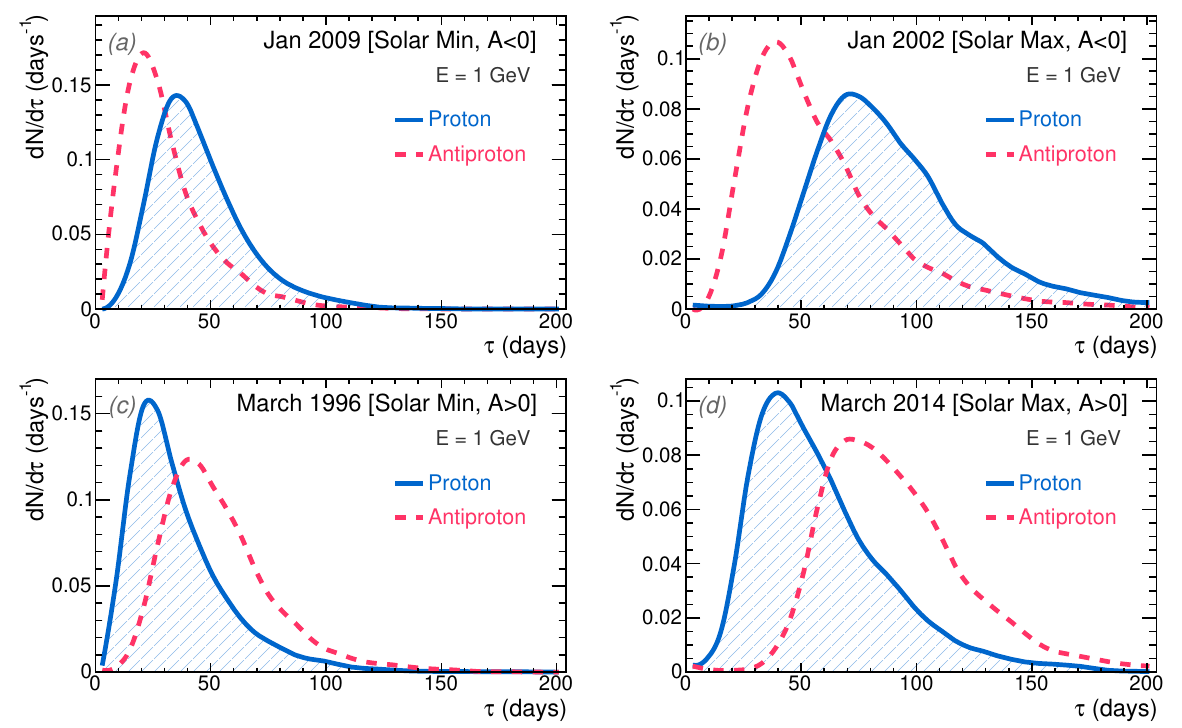}
 \caption{Distribution of propagation times for galactic \GCR protons arrived near the Earth at $1\text{GeV}$ for the $A < 0$ (top panels),
   $A > 0$ (bottom panels) in solar minimum and solar maximum activities (left and right panels, respectively).
   The average propagation times of cosmic protons (antiprotons) in the four epochs are:
   \emph{(a)} $\langle\tau\rangle=$\,45\,days (31\,days),
   \emph{(b)} $\langle\tau\rangle=$\,89\,days (58\,days),
   \emph{(c)} $\langle\tau\rangle=$\,36\,days (54\,days),
   \emph{(d)} $\langle\tau\rangle=$\,59\,days (91\,days).
 }
  \label{Fig::ccHistoTimes}
\end{figure*}

In Fig.\,\ref{Fig::ccHistoTimes}, we present the normalized probability distributions of the \GCR propagation times $\tau$.
For the moment, we focus on cosmic protons with a kinetic energy of $E=$1\,GeV, as indicated by the blue solid lines.
We will discuss cases involving different energies and charge signs later in this discussion.
The plots in the figure are generated by sampling, following the method described in Sect.\,\ref{Sec::SDE},
from the simulation of 15,000 pseudo-proton trajectories.
Each panel corresponds to a distinct reference epoch in the solar polarity cycle.
%
%
%
From the figure, it is evident that the distributions exhibit
pronounced asymmetric tails towards larger $\tau$ values.
The asymmetric nature of these distributions may be attributed, in part, to the effects of drift, as discussed later,
and is partially influenced by the fact that trajectory times are bounded by a lower kinematic limit.
From these distributions, we can readily calculate the average value $\langle\tau\rangle$ and peak values,
following Eq.\,\ref{Eq::AvgTime}, for various solar epochs.
During solar minimum (left panels) for $A<0$, the distribution peaks at 35 days (a), whereas for $A>0$, it peaks at 22 days (c).
During solar maximum (right panels), the distribution for $A<0$ peaks at 71 days (b), and for $A>0$, it peaks at 40 days (d).
Hence, for 1\,GeV protons, propagation times of \GCRs during $A<0$ epochs are generally longer compared to $A>0$ epochs.
Furthermore, propagation times during solar maximum are consistently longer compared to solar minimum,
indicating that the level of solar activity significantly influences \GCR propagation times.
These observations can be explained in terms of drift and diffusion processes.
The effects of drift, particularly those arising from IMF gradient and curvature,
drive charged particles along preferential paths, depending on the IMF polarity state \cite{Potgieter2021}.
During periods of negative IMF polarity, galactic protons reach the inner heliosphere through equatorial regions (colatitude $\vartheta\,{\approx}\,90^\circ$).
In contrast, during positive polarity epochs, they preferentially pass through the polar regions ($\vartheta\,{\approx}\,0^\circ$ and $\vartheta\,{\approx}\,180^\circ$).
This situation is illustrated in Fig.\,\ref{Fig::ccTrajectory}, which depicts two typical trajectories
of pseudo-protons with a kinetic energy of 1\,GeV traveling from the heliopause to the vicinity of Earth.
These trajectories represent opposite IMF polarity scenarios.
Thus, as a result of drift, \GCRs probe different regions of the heliosphere characterized by different plasma conditions.
Polar regions exhibit faster solar wind speeds and weaker IMF strength, leading to increased convection compared to equatorial regions.
However, particles traveling through equatorial regions encounter the heliospheric current sheet,
resulting in a substantial increase in the length and duration of their trajectories.
The combination of these processes leads to different average values for propagation times and energy losses.
As we have found, cosmic protons experience larger propagation times during epochs with $A<0$ polarity.
Additionally, the overall change in the diffusion coefficient over the long term also contributes to variations
in propagation times, with a generally faster propagation during solar minimum compared to solar maximum.
%
%
\begin{figure}
 \centering
\centering\includegraphics[width=0.44\textwidth]{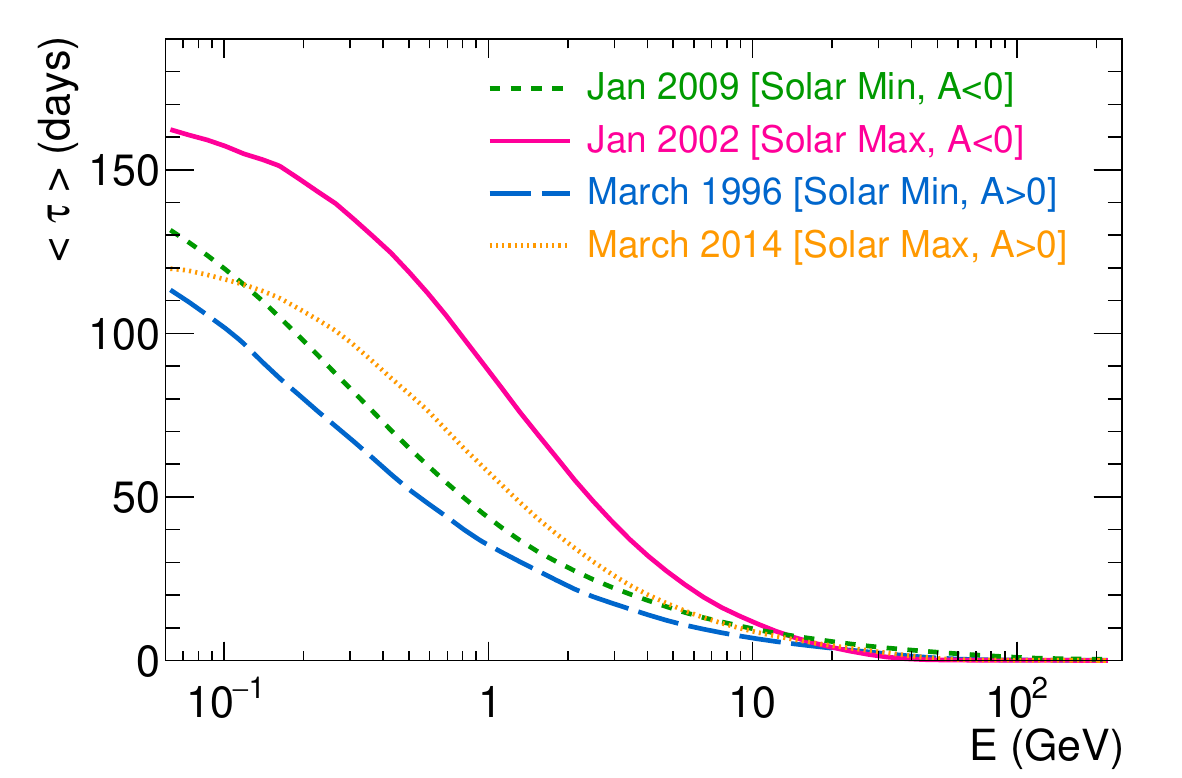} 
\caption{Propagation times versus the energy of the galactic \GCR protons arrived near the Earth
  for the $A < 0$ and $A > 0$ in solar minimum and solar maximum activities.}
  \label{Fig::ccPropTimeVSEnergy} 
\end{figure}

To better understand the situation, it is instructive to examine the relationship between the propagation times of \GCRs and their energy.
In Fig.\,\ref{Fig::ccPropTimeVSEnergy}, we calculate and plot the average propagation times of protons as a function of their arrival kinetic energy $E$.
The figure presents four sets of calculations for $\langle\tau\rangle$ as a function of energy, covering the reference epochs considered and ranging from $50$\,MeV to $250$\,GeV.
Both Fig.\,\ref{Fig::ccHistoTimes} and Fig.\,\ref{Fig::ccPropTimeVSEnergy} reveal intriguing features
reflecting the relationship between propagation time and \GCR transport, which can be understood
through simple physical considerations.
In a diffusion dominated scenario, the propagation time $\tau_{d}$ would exhibit a linear decrease with increasing diffusivity,
given by $\tau_{d}\approx \int_{r_0}^{r_{b}} \frac{r}{3K}dr$. This relationship depends on the diffusion coefficient $K$ and
consequently its energy or rigidity dependence \citep{OGallagher1975}.
In particular, as expressed in Eq.\,\ref{Eq::KvsR}, the rigidity dependence of \GCR diffusion
can be characterized as $K(R)\sim \beta R^{\eta}$, where the spectral index takes the value $\eta=a$
for $R \ll R_{0}$ and $\eta=b$ for $R \gg R_{0}$.
In practice, based on the reconstruction presented in Fig.\,\ref{Fig::ccTransportParameters},
the indices $a$ and $b$ are similar to each other and their variations are relatively modest.
Therefore, to first approximation, one can consider $a \approx b \cong \eta$. 
Consequently, the diffusion time can be expressed as $\tau_{d} \propto (K_{0} \beta R^{\eta})^{-1}$,
with $\eta\approx$\,1.
The temporal dependence of \GCR diffusion is primarily captured by the parameter $K_{0}(t)$ \citep{Tomassetti2018PHeVSTime}.
Its relationship with solar activity can be elucidated with the plots of
Fig.\,\ref{Fig::ccTransportParameters} (i) and (iv), 
if we focus on the timescales of several months, where short-term features such as double-peak maxima (in SSN) or Forbush decreases (in GCRs) can be ignored.  
In this long-term behavior, 
the diffusion coefficient appears anti-correlated with the SSN cycle. 
Thus the diffusion time $\tau_{d} \propto K^{-1}_{0}$ must be positively correlated with it. 
During epochs of solar maximum (minimum), \GCR diffusion is slower (faster), resulting in larger (smaller)
propagation times associated with diffusion.
At lower rigidities, the effects of convection become increasingly significant.
Due to the outward direction of the solar wind, convection reduces the likelihood of Galactic particles reaching Earth.
The influence of convection establishes an upper limit on the propagation time,
denoted as $\tau_{c}$, which can be approximated as $\tau_{c} \approx \int_{r_{0}}^{r_{b}} V_{sw}^{-1}dr$ and
is on the order of approximately $\sim 500$\,days.
Conversely, a lower bound of the order of $\sim$\,1\,day can be obtained as ballistic limit,
given by $\tau_{b} = (r_{b}-r_{0})\beta^{-1}c^{-1}$ for rectilinear \GCR propagation.
Based on these considerations, in a diffusion-convection scenario, the resulting propagation
time can be derived as a combination of $\tau_{d}$ and $\tau_{c}$ \citep{Strauss2011JGR}:
\begin{equation}
  \tau_{cd} \approx \left( \tau_{d}^{-1} + \tau_{c}^{-1} \right)^{-1} 
\end{equation}
By using typical values such as $V_{sw}\approx\,450$\,km\,s$^-1$, $K_{0}\approx 10^{23}$\,cm\,s$^{-2}$, and $r_{b}\approx 120$\,AU,
one would anticipate that the propagation time $\tau_{cd}$ reaches approximately 6 months in the sub-GeV energy region,
where convection dominates.
This would be followed by a decrease in the form of $\tau\propto E^{-\eta}$ in the multi-GeV region, where diffusion becomes the dominant process.
These considerations provide an explanation for the general characteristics observed in Fig.\,\ref{Fig::ccPropTimeVSEnergy}.
%
\begin{figure*}
\centering
\centering\includegraphics[width=0.8\textwidth]{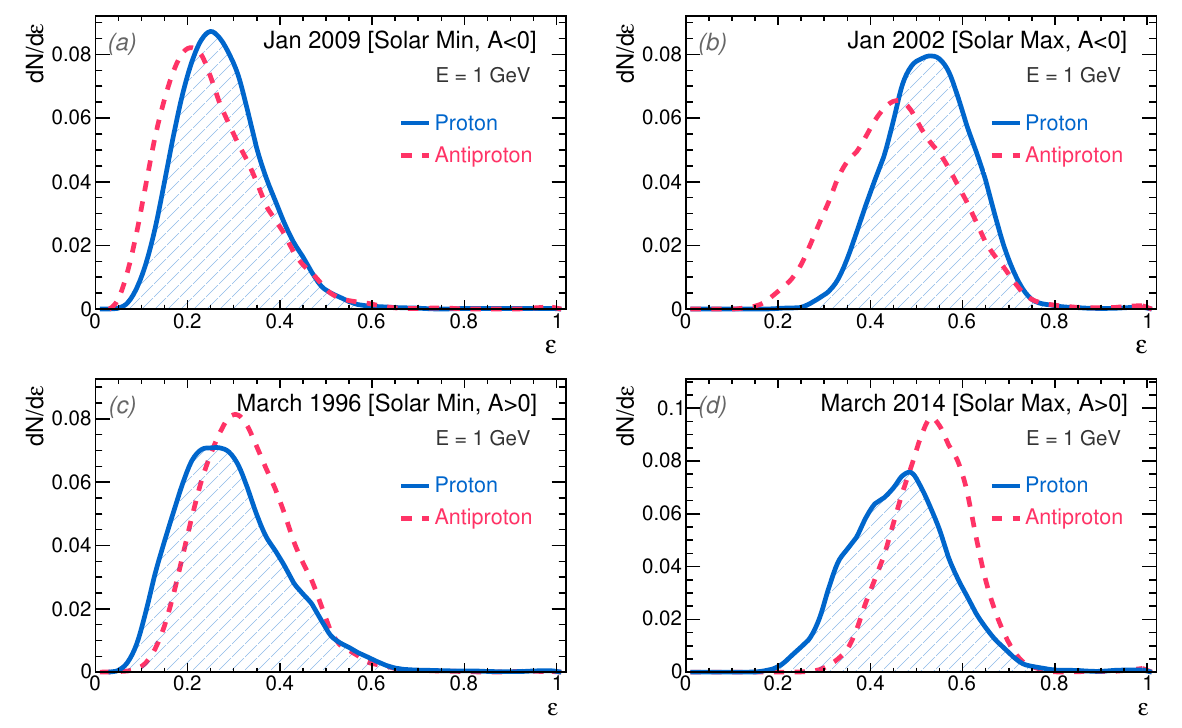}
\caption{Distribution of the energy loss fraction $\varepsilon$ for \GCR protons arrived near the Earth at $1\text{GeV}$ for
  the $A < 0$ (top panels), $A > 0$ (bottom panels) in solar minimum and solar maximum activities (left and right panels, respectively).
  The average loss fractions of cosmic protons (antiprotons) in the four epochs are:
  \emph{(a)} $\langle\varepsilon\rangle=$\,0.28\, (0.25),
  \emph{(b)} $\langle\varepsilon\rangle=$\,0.53\, (0.47),
  \emph{(c)} $\langle\varepsilon\rangle=$\,0.30\, (0.33),
  \emph{(d)} $\langle\varepsilon\rangle=$\,0.47\, (0.53).
  }
  \label{Fig::ccHistELoss}
\end{figure*}
%
%
To explain the differences between the various curves, it is essential to consider polarity-dependent effects, 
although drift effects can only be evaluated numerically.
In contrast to a diffusion-dominated scenario, the influence of drift is to reduce the propagation time
in both polarity states, which may appear counterintuitive.
A clear explanation for this effect is provided in \citet{Strauss2011JGR}.
The action of drift introduces preferential trajectories for \GCRs reaching Earth while suppressing non-preferential ones.
In many instances, this selection process favors straighter trajectories, corresponding to smaller propagation times.
This aspect contributes to the observed asymmetric tails in the distributions and helps explain the behavior observed
in Fig.\,\ref{Fig::ccPropTimeVSEnergy}.

An interesting feature that stands out in Fig.\,\ref{Fig::ccPropTimeVSEnergy} is the crossing point between curves (d) and (b),
occurring at kinetic energies below $\sim 100$\,MeV. This crossing point is located between the yellow dotted line
(representing solar maximum, $A>0$) and the green short-dashed line (representing solar minimum, $A<0$).
For high-energy \GCRs, the most dominant factor affecting the propagation times is the level of solar activity.
In this energy regime, the mean propagation times during solar maximum are consistently larger than those during
solar minimum, as observed in Fig.\,\ref{Fig::ccHistoTimes} for $E=$1 GeV.
However, below the crossing energy, the polarity effect becomes dominant, with mean propagation times during negative
polarity cycles being consistently larger than those during positive cycles.
A possible explanation for this behavior may involve the interplay between diffusion and drift at the different rigidities.
However, it is important to note that the two maxima (b) and (d) belong to different solar cycles (23 and 24),
which can vary in terms of intensity and turbulence properties.
From a look at Fig.\,\ref{Fig::ccTransportParameters}, this is indeed the case. 
For instance, the maximum of cycle 24 (d) was the smallest recorded in a century of SSN observations and exhibited a double-peak structure.
Similarly, the two minima (a) and (c) belong to different solar cycles, with the 2009 minimum across cycles 23/24 (a) being unusually
weak and long compared to other minima. As shown in Fig.\,\ref{Fig::ccTransportParameters}, these different minima and maxima
of recent solar cycles result in variations in the \GCR transport parameters, both in terms of normalization and slope.

\subsection{Energy losses} 
\label{Sec::EnergyLosses}  

We now turn our attention to the energy losses. In particular, we focus on the loss fraction $\varepsilon=\Delta E/E_{0}$.
In Fig.\,\ref{Fig::ccHistELoss}, the probability distributions of $\varepsilon$ are shown for cosmic protons (solid blue lines)
reaching Earth with a kinetic energy of $E=1$\,GeV.
%
\begin{figure}
 \centering
\centering\includegraphics[width=0.44\textwidth]{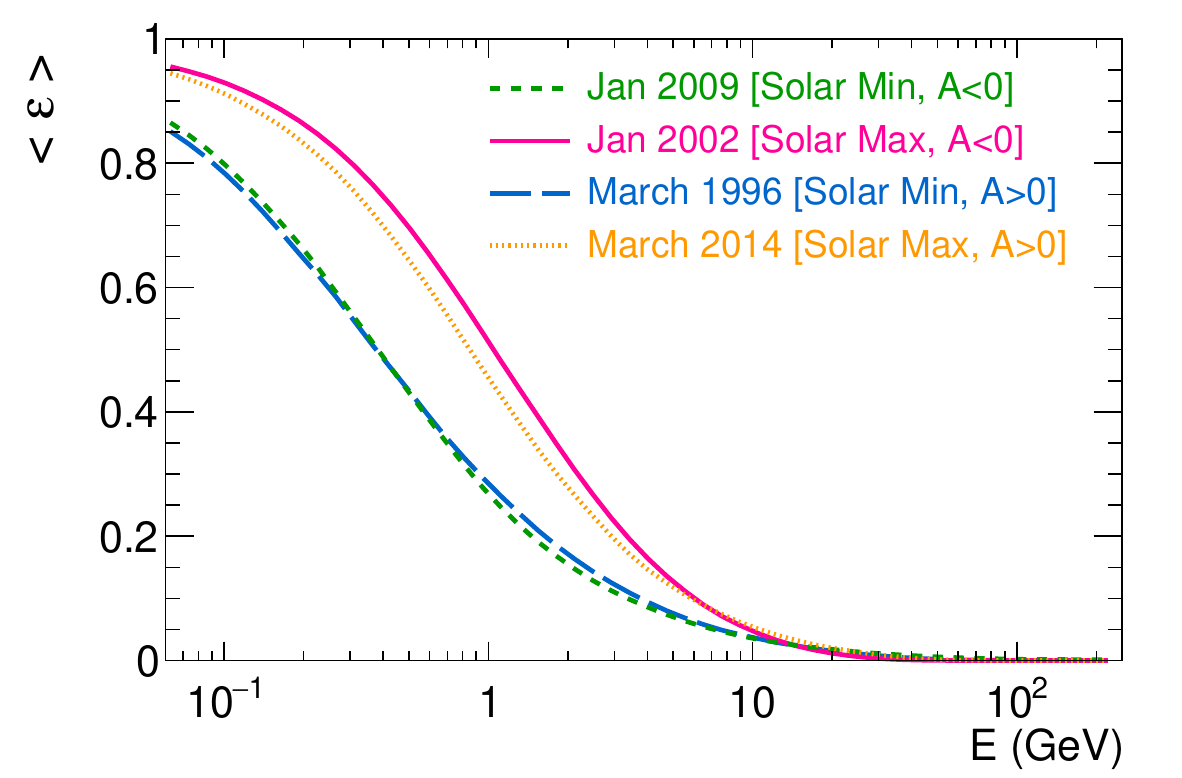} 
\caption{Energy loss rate $\varepsilon=\Delta E/E_{0}$ as a function of the near-Earth
  kinetic energy $E$ of \GCR protons in four reference epochs of the solar cycle}.
  \label{Fig::ccELossVSEnergy}
\end{figure}
%
%
The calculation method employed is similar to that of the propagation times, involving the Monte-Carlo sampling 
through the simulation of 15,000 pseudo-particle trajectories at a fixed epoch, energy, and particle type.
The panels in the figure correspond to the same four reference epochs as discussed in the previous section,
representing different solar cycle intensities and polarities.
In comparison with the $\tau$ distributions, the $\varepsilon$ distributions exhibit less skewness, 
although asymmetric tails can still be observed in panels (a) and (c), which correspond to the 2009 and 1996 solar minima, respectively.
During solar minimum, the average fraction is $\langle\varepsilon\rangle\sim$\,0.3:
protons detected near Earth with kinetic energy of 1\,GeV have
lost nearly one third of their initial kinetic energy, on average.
During solar maximum the losses are more significant, reaching values of about 0.4 - 0.5 in panels (b) and (d).
This implies that 1\,GeV protons detected during solar maximum, entered the heliosphere with an average
initial energy of 2\,GeV.
%
%
\begin{figure*}
 \centering
 \centering\includegraphics[width=0.8\textwidth]{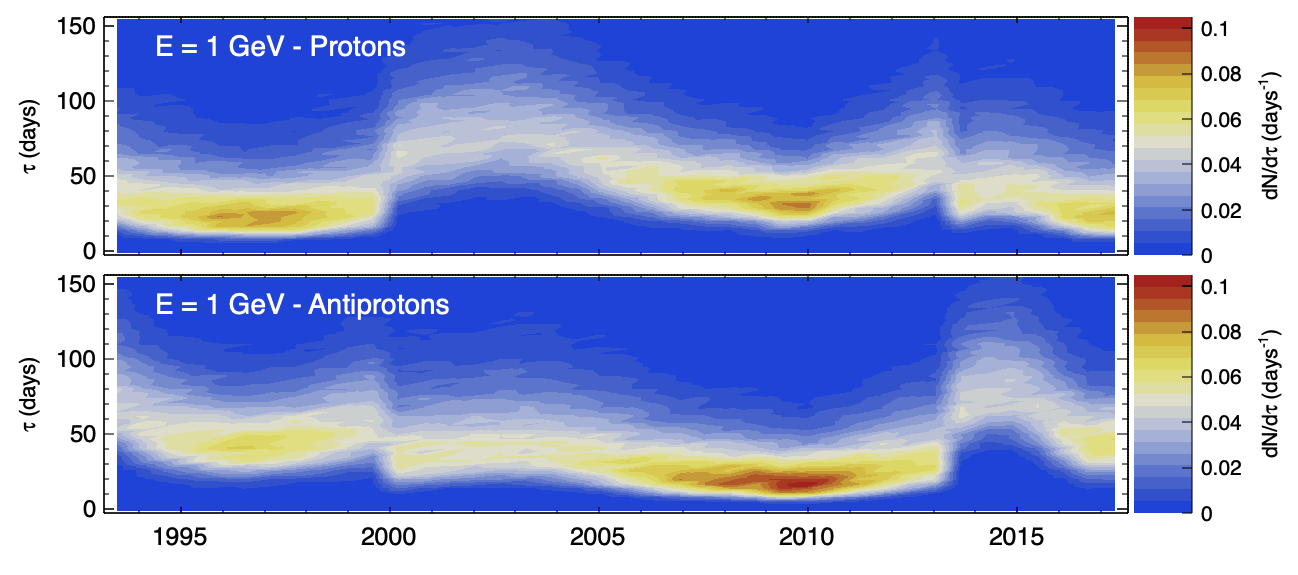} 
 \caption{Temporal evolution of the probability distributions of the propagation time $\tau$ for \GCR protons
   (top) and antiprotons (bottom) reaching Earth with $E=1$\,GeV of kinetic energy.}
  \label{Fig::ccPropTimeVSEpoch2D_PbarP}
\end{figure*}

The dependence of the average loss fraction $\langle\varepsilon\rangle$ on the near-Earth energy $E$ of the protons
is shown in Fig.\,\ref{Fig::ccELossVSEnergy}.
The four curves in the figure correspond to the reference epochs considered in our analysis.
The average loss fractions $\langle\varepsilon\rangle$ are calculated by constructing profile histograms
for the probability distribution as a function of $E$.
It is important to emphasize that, as per the physical mechanisms incorporated in the Parker equation
discussed in Sect.\,\ref{Sec::CRPropagation}, energy changes for cosmic protons consist solely of adiabatic losses.
Consequently, the loss fraction $\varepsilon$ is always positive.
From Fig.\,\ref{Fig::ccHistELoss} and Fig.\,\ref{Fig::ccELossVSEnergy}, some observations can be made.
The loss fraction decreases rapidly with increasing particle energy.
However, at low energies the loss fraction saturate to unity.
Here the dynamics of low-energy particles are primarily governed by convection with the solar wind.
These particles are decelerated and then thermalized before reaching the inner heliosphere.
In contrast, in the high-energy limit the loss fraction becomes negligible.
Comparing the different epochs, it can be noticed that the curves fall into two distinct groups based
on the intensity of solar activity, and they are less sensitive to the polarity states.
In particular, the curves associated with solar minima (a) and (c) are nearly indistinguishable.
The observed behavior of energy losses at different energies and epochs
can be discussed qualitatively in the context of the force-field approximation.
In this approximation, energy losses are captured by the modulation potential $\phi$,
which can be expressed as $\phi \sim \frac{V_{sw}}{3K_{0}}(r_{b}-r_{0})$.
Hence, a qualitative explanation can be provided in terms of the behavior of the $V_{sw}/K_{0}$ ratio.
Notably, from Fig.\,\ref{Fig::ccTransportParameters} (a) and (d), it can be observed that epochs
of solar maximum correspond to slower diffusion, implying smaller values of $K_{0}$, and vice versa.
The solar wind speed $V_{sw}$, on the other hand, does not exhibit a clear long-term trend over the solar cycle.
However, it does display a distinct latitudinal profile, as discussed in Sect.\,\ref{Sec::Heliosphere}.
As previously discussed, the action of drift is to guide \GCR protons along preferential trajectories.
During $A>0$ epochs, protons tend to reach Earth predominantly from the polar region
through the fast wind region.  
During $A<0$ epochs, protons tend to travel predominantly through
the equatorial region of the heliosphere, in the slow wind region.
Based on this observation, one might initially expect higher energy losses during $A>0$ epochs.
However, an opposite effect comes into play due to the presence of the heliospheric current sheet.
Particles reaching Earth during $A<0$ epochs are more likely to drift across the current sheet,
which significantly increases both the total length and duration of their trajectories.
The larger trajectory lengths imply higher energy losses for particles during $A<0$ epochs.
The interplay between these two competing effects, convection and heliospheric current sheet,
results in a nuanced relationship between energy loss and IMF polarity.
From Fig.\,\ref{Fig::ccELossVSEnergy}, it can be noted that the curves representing
opposite polarity states intersect between kinetic energies of 3 and 10 GeV.
This crossing feature likely reflects the interplay of these effects.

\subsection{Particles and antiparticles} 
\label{Sec::Antiparticles}               

In the previous paragraphs, the discussion was primarily centered around \GCR protons.
We now analyze the results for antiprotons.
For antiprotons, all the relevant calculations were conducted using the same SDE model
discussed in Sect.\,\ref{Sec::CRPropagation}.
The key parameters of the model were those determined from \GCR proton data as described in Sect.\,\ref{Sec::TransportParameters}.
Such approach relies in the assumption that the key parameters governing the transport
of protons can describe other nuclear species with different mass, charge or charge-sign.
For diffusion, the \GCR mean free paths are universally determined by the particles rigidity \citep{Tomassetti2018PHeVSTime,Tomassetti2019Numerical}.
Moreover, both diffusion or convection are inherently insensitive to the charge-sign of the particles.
In fact, from a microscopic point of view, both effects arise from the scattering of charged particles
off small-scale irregularities of the IMF, for which there is no distinction between protons and antiprotons.
This is not the case for drift processes across the large-scale component of the IMF.
The drift speed of particles depends explicitly on the product ${q}A$, where $\hat{q}=q/|q|$ represents the charge sign, and $A$ is the polarity of the IMF.

In our model, the $\hat{q}A$ symmetry is rigorously respected. Consequently, a single proton under $A>0$ polarity
experiences the same dynamics as an antiproton under $A<0$ polarity,
assuming that all other kinematics and plasma conditions are identical.
When our description pertains to the differential flux as a function of energy near Earth, however,
minor differences may arise between protons and antiprotons due to their different LIS shapes,
as illustrated in Fig.\,\ref{Fig::ccLISProtonAntiproton}.
In fact we note that, even if we have often considered particles at some fixed energy (\eg, $E=1$\,GeV),
that energy represents the \GCR final energy which results from the solar modulation process.
This implies a convolution over initial LIS of kinetic energies $E_{0}$.
Furthermore, it is crucial to emphasize that in this work, the calculations are performed for
actual solar cycles. When comparing plots associated with $A<0$ and $A>0$ (\eg, January 2009 and March 1996), 
it is  important to recognize that we are not considering the sole effect of a simple polarity switch.
Instead, we are examining two distinct solar cycles with opposite polarities, where various other parameters can also differ.

A comparison of proton and antiproton propagation times is provided in Fig.\,\ref{Fig::ccHistoTimes}
for kinetic energies of $E=1$\,GeV. In the figure, the distributions of antiprotons are represented by pink dashed lines.
Observing panels (a) and (b), it can be seen that during both epochs of negative polarity, the distributions of
antiproton propagation times peak at lower values compared to protons.
In other words, in $A<0$ epochs, antiprotons propagate faster than protons.
Conversely, in panels (c) and (d), the opposite scenario is evident for $A>0$ epochs.
During these periods, the antiproton propagation times are significantly larger than those of protons.
When comparing the two situations, namely the top panel with the bottom panels, the $\hat{q}A$ symmetry effect becomes evident.
The act of polarity reversal seems to effectively \emph{exchange} the distributions of protons and antiprotons,
so that the propagation times depend mainly on the sign product $\hat{q}A$.  
Explanations for these differences can be attributed to the polarity-dependent action of drift
in the trajectories of \GCRs, as discussed in Sect.\,\ref{Sec::PropagationTimes}.
In fact, the two typical trajectories illustrated in Fig.\,\ref{Fig::ccTrajectory} were calculated in terms of $\hat{q}A$.
This means that the trajectory of a proton under positive polarity is identical to that of
an antiproton under negative polarity, assuming other kinematics or heliospheric factors remain the same.
It is also interesting to compare the evolution of the probability distributions over the entire solar cycle.
In Fig.\,\ref{Fig::ccPropTimeVSEpoch2D_PbarP}, the normalized probabilities of \GCR protons (top)
and antiprotons (bottom) at $E=1$\,GeV are displayed as a function of the epoch for the considered time interval.
In these plots, the probability is represented on the $z-$axis and is represented by the color palette on the right
side. Particularly, the yellow-red serpentine pattern illustrates the temporal evolution of the probability peak.
From the figure, it can be noticed that the IMF reversals occurred in 2000 and in 2013 had a significant impact,
causing two shifts in the distributions.

During the 2000 reversal, when passing from $A>0$ to $A<0$ polarity, the proton distribution exhibited an
increase to larger propagation times, while in the 2013 reversal, the shift was in the opposite direction.
On average, the mean propagation time of \GCR protons under positive polarities is between two and three months, roughly,
while during positive polarity it is about one month.
For antiprotons, the effect of magnetic reversal on the $\tau$ distributions is also noticeable, leading to similar shifts but in opposite directions.
In addition to the effect of magnetic reversal, it is interesting to examine the distributions during a consistent polarity state.
For instance, between 2000 and 2013, the distributions of propagation times for \GCR protons and antiprotons
follow a similar evolution over the solar cycle.
This suggests that, as long the polarity is well defined and does not change, even when drift processes are at play,
the temporal dependence of propagation times is primarily driven by diffusion and convection.
\begin{figure*}
 \centering
 \centering\includegraphics[width=0.8\textwidth]{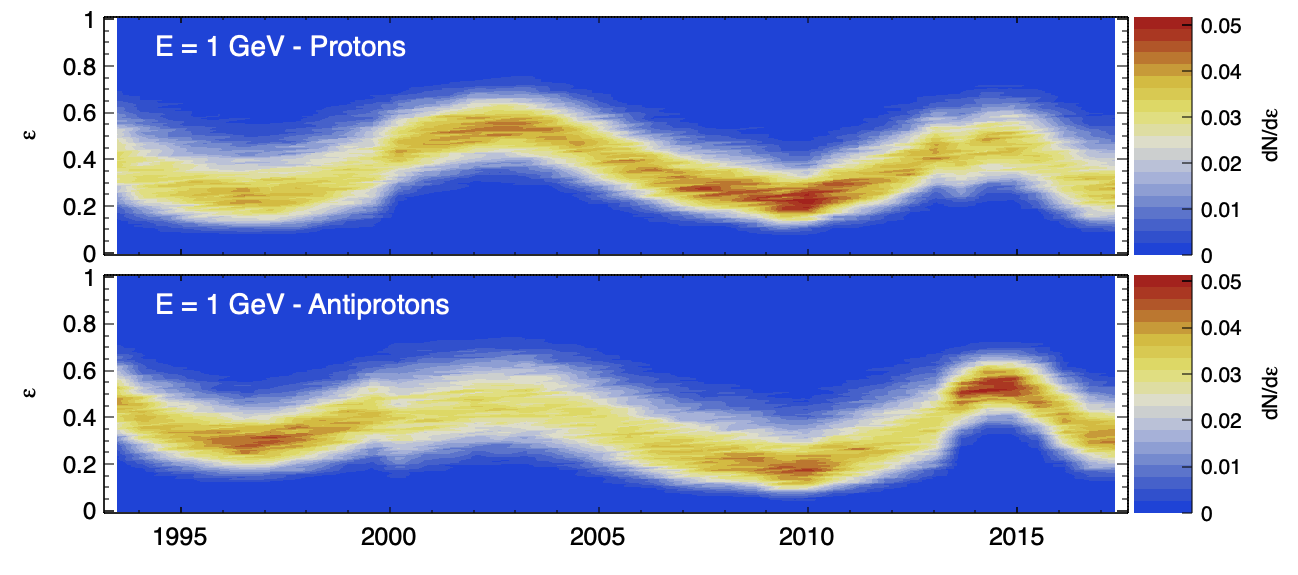} 
 \caption{Temporal evolution of the probability distributions of the loss fraction $\varepsilon$
   for \GCR protons (top) and antiprotons (bottom)
   reaching Earth with $E=1$\,GeV of kinetic energy.}
  \label{Fig::ccELossVSEpoch2D_PbarP}
\end{figure*}

Now, we shift our focus to the energy losses. The probability distributions of the loss fraction $\varepsilon$
are shown in Fig.\,\ref{Fig::ccHistELoss} for particles with $E=1$\,GeV.
Once again, the distributions of antiprotons are denoted by pink dashed lines.
Again, we observe some differences between antiprotons and protons,
including the presence of a $\hat{q}A$-symmetry if we compare the top panels (a, b) with the bottom panels (c, d).
However, it is important to note that distinctions in energy losses between protons
and antiprotons are not as pronounced as those observed in the case of propagation times.
As we explored in Sect.\,\ref{Sec::EnergyLosses},
the loss fraction seems to be more responsive to the intensity of solar activity rather than to the IMF polarity.
This characteristic is also evident in Fig.\,\ref{Fig::ccELossVSEpoch2D_PbarP}, showing the long-term evolution of the
normalized probability distributions of $\varepsilon$.
The figure illustrates that during the two IMF reversals in 2000 and 2013, there is only a modest shift in the distributions.
Conversely, a significant evolution over the solar cycle is evident.
The yellow-red serpentine pattern seems to follow a nearly continuous trend, resembling a quasi-sinusoidal shape.
This overall dependence over the solar cycle can be interpreted in terms of adiabatic losses, as discussed
in Sect.\,\ref{Sec::EnergyLosses}, since the loss process is charge-sign blind and makes no distinction
between protons and antiprotons.
Nevertheless, these processes are coupled with drift, which impacts the trajectories of \GCRs within the large-scale IMF.
From one side, the $\hat{q}A$-dependent drift cause \GCRs of positive (negative) charge to travel preferentially
through the fast-wind regions of the heliosphere during $A>0$ ($A<0$) polarity epochs. 
In the fast-wind regions, particles are subjected to larger loss rate in comparison to particles of the slow-wind region.
On the other hand, particles in the polar (fast-wind) regions make straighter trajectories and with smaller propagation times,
in comparison with particles passing across the equatorial (slow-wind) region. In fact the latter
are influenced by the heliospheric current sheet, which cause longer trajectories and larger propagation times,
which implies more severe energy losses.

The absence of pronounced shifts in the $\varepsilon$ distributions across the IMF reversal
may be attributed to the interplay of two competing effects, as discussed in Sect.\,\ref{Sec::EnergyLosses}.
On one hand, the $\hat{q}A$-dependent drift leads \GCRs with positive (negative) charge to preferentially
travel through different wind regions where cosmic particles are exposed to different loss rates.
On the other hand, particles coming from the polar regions follow shorter trajectories and with smaller losses,
in comparison to particles passing through the equatorial regions, due to the presence of the current sheet.
As for the mean propagation times $\langle\tau\rangle$ and the loss fraction $\langle\varepsilon\rangle$,
their energy dependence for \GCR antiprotons exhibits a qualitative similarity to those of protons,
as shown in Fig.\,\ref{Fig::ccPropTimeVSEnergy} and Fig.\,\ref{Fig::ccELossVSEnergy}.
These opposing effects contribute to the relatively stable distributions for the loss factors across IMF reversals.

\section{Summary and Conclusions} 
\label{Sec::Conclusions}          

After entering the heliosphere, \GCRs spend several days traveling the interplanetary plasma
before reaching our detectors near Earth.
During this journey, \GCRs are influenced by fundamental physical mechanisms which include diffusion, drift, convection,
and adiabatic cooling.
The interplay among these mechanisms depends on the turbulence conditions of the background plasma, which are influenced by the Sun's magnetic activity.
In this paper, we have conducted calculations to assess the propagation times energy losses of \GCRs in the heliosphere.
While $\varepsilon$ and $\tau$ are not directly measurable quantities, we have devised a data-driven approach
to obtain their probability distributions within the framework of a solar modulation model that has been
globally calibrated through \GCR proton measurements.
To determine the evolution of the model parameters illustrated in Fig.\,\ref{Fig::ccTransportParameters},
we employed a data-driven approach which 
relies on conventional techniques of statistical inference \citep{Fiandrini2021,Tomassetti2023MFP}.
The transport parameters were constrained using 25 years of data, consisting of energy- and time-resolved measurements of \GCR
protons from AMS-02, PAMELA, EPHIN/SOHO, BESS, and Voyager-1.
Additionally, we incorporated essential inputs in the form of heliospheric parameters,
as shown in Fig.\,\ref{Fig::HeliosphericParameters}.
These parameters are used to capture the varying conditions of the IMF.
They were derived from data collected by spacecraft such as WIND and ACE, as well as from the Wilcox Solar Observatory.
Using our proton-calibrated model, we computed propagation times and energy losses through trajectory-based sampling.
This allowed us to examine the shapes of probability distributions, analyze their dependence on the particle energy,
how they are influenced by IMF polarity reversals, and track their evolution throughout the solar activity cycle

Our solar modulation framework relies on the SDE method. From a computational standpoint, our approach is similar
to recent works that also employ the SDE method \citep{Strauss2011JGR,Strauss2013,Vogt2022}.
The primary goal of this work, however, was not solely the computation of $\tau$ and $\varepsilon$ distributions.
Rather, we focused on investigating their evolution through the actual solar cycles.
We can compare our results for the  mean propagation time with the work of \citet{Moloto2023}
which is based on similar calculations but completely different data.
Their results appear in good agreement with determination of $\langle\tau\rangle$.

Our results may have implications for understanding the dynamics of the \GCR modulation phenomenon,
particularly concerning the association between \GCR transport mechanisms and solar variability.
Many studies on this subject have explored the time lag between the \GCR fluxes and  solar activity,
often attributed to the dynamics of the expanding solar wind \citep{Hathaway2015,Usoskin1998,Tomassetti2017TimeLag}.
However, recent studies have reported a 22-year quasi-periodic variation in the lag, following the IMF polarity cycle,
and have observed a significant dependence on the particle energies \citep{Tomassetti2022PRD,Koldobskiy2022TimeLag,Shen2020TimeLag}.
These works suggested that this feature is due to the energy-dependent propagation times of \GCRs in the
heliosphere, which is well supported by our findings.
Specifically, it has been suggested that the \GCR modulation lag comprises two components:
an energy-independent lag of approximately 4-16 months from the expansion of the background plasma,
and an energy-dependent lag that reflects \GCR propagation times\,\citep{Tomassetti2017TimeLag}.
The investigation of the \GCR propagation times may be of help for advancing the development of 4D time-dependent models of \GCR modulation.
A significant challenge in this repect lies in accurately describing the dynamics of \GCRs as they traverse a variable
medium \citep{Engelbrecht2022,FlorinskiPogorelov2009,Pei2010}.
Many effective models, including the framework we employed, rely on a quasi-steady state approximation
where the temporal evolution of \GCR modulation is represented as a continuous sequence of equilibrium solutions.
To rely in this approximation, it is essential to have a comprehensive understanding of the
timescales associated with \GCR propagation in relation to the changing solar activity.
Therefore, having estimates of \GCR propagation times as a function of their energy and
for different phases of the solar cycle is crucial.

The study presented here has several limitations that we briefly discuss.
One limitation lies in the minimal number of parameters to capture the variable conditions of interplanetary plasma.
Expanding the model to incorporate more observations and proxies could provide
a more accurate representation of the heliospheric environment.
Examples are the account of variations in solar wind speed or changes in the termination shock \citep{Manuel2014}.
Similarly, to constrain the \GCR transport, the account of other time-dependent parameters should be considered.
One example is our assumption of the same parameters describing parallel and perpendicular diffusion,
while other works observed different dependencies \citep{Corti2019}.
Another example is the drift parameters $K_{A}^{0}$ or $R_{A}$ of Eq.\,\ref{Eq::KA}, which we have set to constant values over the solar cycle.
However, one may argue that drift may be suppressed at large tilt angles, and in particular in the reversal phase.
The role of drift over the solar cycle can be investigated using the new 
time-dependent measurements on GCR antiprotons from AMS-02. will be precious to study
Regarding energy changes, we have assumed that only adiabatic cooling is at work.
Nonetheless, at energies lower that those considered in our analysis, reacceleration processes of \GCRs
at the termination shock may influence the evaluation of the loss faction $\varepsilon$ and its probability distribution.
Moreover, in the outer region beyond the termination shock, the cooling is less effective and the loss fraction may
be even affected by regions with adiabatic heating \citep{Langner2006}.
Moreover, we have provide a two-dimensional spatial description of the \GCR propagation, assuming azimuthal symmetry
for the IMF and a spherically shaped structure for the heliopsheric boundaries (\ie, the heliopause and the termination shock). 
The large values of the propagation times indicates that \GCRs observed near Earth
sample large volumes of the heliosphere, including the heliosheat and the heliotail.
While the \GCR transport in the heliosheat is incorporated in our model,
the account for an extended heliotail is missing, as it requires the implementation of a non-spherical model for the heliopause \cite{Boschini2019}. 
One may expect that the existence of such an extended modulation region may produce larger tails in the distributions of the GCR propagation times. 
In fact, the population of GCRs propagating inside the region of the heliotail would spend significant time in that region,
before eventually reaching the inner heliosphere or the interstellar space.
On the other hand, although past models of the heliosphere broadly assumed an magnetosphere-like shape with an extended heliotail,
recent observations from space probes such as Voyager, Cassini and IBEX have disfavored such a model,
suggesting a bubble-like heliosphere \cite{Dialynas2017,Zhang2024}. 
Nonetheless, the account of a truly 3D formulation would enable the model to capture effects related to the
detailed large-scale structures of IMF and solar wind. 
Overcoming all these limitations could enhance the model ability to capture the complexity of
the problem, possibly leading to a more comprehensive understanding of the solar modulation phenomenon.

\section*{Acknowledgements} 

We acknowledge the support of Italian Space Agency (ASI) under agreement ASI-UniPG 2019-2-HH.0. 
B.K. acknowledges support from agreement ASI-INFN 2019-19-HH.0.
N.T. acknowledges the project \emph{SMILE - Solar ModulatIon of Light Elements in cosmic rays},
in the program Fondo Ricerca 2021 of the University of Perugia, Italy.
The \GCR data were retrieved from the ASI-SSDC \emph{\href{https://tools.asdc.asi.it/CosmicRays}{Cosmic Ray Data Base}}.
Spacecraft data were extracted from the \emph{\href{https://omniweb.gsfc.nasa.gov}{OMNIWeb}} service of the NASA Space Physics Data Facility.
Solar data were taken from the Wilcox Solar Observatory at Stanford University and from the
 \emph{\href{https://www.sidc.be/silso/}{SILSO}} SSN database of the \emph{Solar Influences Data Analysis Center} at the Royal Observatory of Belgium.


\end{document}